\documentclass[runningheads]{llncs}
\usepackage[T1]{fontenc}
\usepackage{graphicx}
\usepackage{amsmath}
\usepackage{amssymb}
\usepackage{mathtools}
\usepackage{multirow}
\usepackage{cite}
\usepackage{hyperref}
\usepackage[table,xcdraw]{xcolor}
\usepackage{pifont}
\usepackage{enumitem} 
\usepackage[]{algorithm2e}

\DeclareMathOperator{\E}{\mathbb{E}}

\SetKwInput{kwCompute}{Compute}
\SetKwInput{kwDefine}{Define}
\SetKwInput{kwInput}{Input}

\usepackage{color}
\begin{document}
\title{Non-Redundant Combination of Hand-Crafted and Deep Learning Radiomics: Application to the Early Detection of Pancreatic Cancer}

\titlerunning{Non-Redundant Combination of Hand-Crafted and Deep Learning Radiomics}
%
\author{Rebeca V\'etil\inst{1,2} \and
Cl\'ement Abi-Nader\inst{2} \and
Alexandre B\^one\inst{2} \and
Marie-Pierre Vullierme\inst{3} \and
Marc-Michel Roh\'e\inst{2} \and
Pietro Gori\inst{1} \and
Isabelle Bloch\inst{1,4}
}


%
\authorrunning{V\'etil at al.}

\institute{LTCI, T\'el\'ecom Paris, Institut Polytechnique de Paris, France \and
Guerbet Research, Villepinte, France \and
Department of Radiology, Hospital of Annecy-Genevois, Universit\'e de Paris, France \and
Sorbonne Universit\'e, CNRS, LIP6, Paris, France \\
\email{rebeca.vetil@guerbet.com}}
\maketitle              

\begin{abstract}We address the problem of learning Deep Learning Radiomics (DLR) that are not redundant with Hand-Crafted Radiomics (HCR). To do so, we extract DLR features using a VAE while enforcing their independence with HCR features by minimizing their mutual information. The resulting DLR features can be combined with hand-crafted ones and leveraged by a classifier to predict early markers of cancer. We illustrate our method on four early markers of pancreatic cancer and validate it on a large independent test set. Our results highlight the value of combining non-redundant DLR and HCR features, as evidenced by an improvement in the Area Under the Curve compared to baseline methods that do not address redundancy or solely rely on HCR features. 

\end{abstract}
\section{Introduction}
Computational methods in medical imaging hold the potential to support radiologists in the early diagnosis of cancer, either by detecting small-size abnormal neoplasms~\cite{litjens2017survey}, or even earlier in the disease course by recognizing indirect signs of malignancy. Such signs are usually subtle and organ-dependent, thus requiring a time-consuming and demanding clinical assessment. For example, in the case of pancreatic cancer, radiologists analyze the overall shape of the organ, check for fat replacement and note whether the pancreas shows atrophy and/or senile characteristics~\cite{matsuda2019age, miura2020focal, khoury2017clinical}. 
The identification of cancerous signs using automated tools can be based on radiomics, which are descriptors of texture and shape of a medical image, computed based on spatial relationships between voxels and their intensity distribution~\cite{kumar2012radiomics, lambin2012radiomics}. Radiomics can be divided into two categories: (i) Hand-Crafted Radiomics (HCR), which are based on predefined mathematical formulas~\cite{kumar2012radiomics, lambin2012radiomics}; (ii) Deep Learning Radiomics (DLR), estimated using deep neural networks~\cite{shafiee2017discovery,kumar2015lung}, which may unveil additional complex relationships between voxels. HCR are generally extracted by open-source frameworks such as pyradiomics~\cite{van2017computational}. While such tools facilitate the standardization of the HCR, they only provide a limited number of predefined features. On the other hand, DLR features are typically extracted using either discriminative or generative models. Discriminative models frequently rely on one or multiple simple CNNs~\cite{paul2016deep,lao2017deep,chen2016automatic,antropova2017deep,huynh2016digital}. To prevent overfitting, some methods extract DLR by utilizing pretrained models trained on large datasets like ImageNet~\cite{paul2016deep,huynh2016digital,antropova2017deep}. The deep neural networks commonly employed for computing these DLR features consist of multiple layers, with each layer producing potential features as its output. As a result, the choice of the layers to retain varies, with each method employing different heuristics to identify them~\cite{paul2016deep, huynh2016digital}. In the realm of generative models, auto-encoder (AE) networks are widely used~\cite{afshar2019handcrafted}. AEs encode an image in a latent vector that is subsequently used to reconstruct the original image. This latent vector is considered to encapsulate the most descriptive features of the input image, making it a natural choice for representing the DLR~\cite{ravi2016deep, kumar2015lung}.

The two types of radiomics are complementary: the computation of DLR is data-driven, which ensures that the extracted features are adapted to a specific problem or type of data. On the other hand, the predefined and generic definitions of HCR may make them less adapted for a given specific task, but favors generalization and interpretability. Therefore, it has been recently proposed to combine HCR with DLR, arguing that this approach could result in an improved feature set for predictive or prognostic models~\cite{afshar2019handcrafted}. The literature reports two main approaches to perform this combination: decision-level methods that train separate classifiers on DLR and HCR before aggregating their predictions \cite{huynh2016digital, antropova2017deep, liu2017pulmonary}, and feature-level methods that concatenate the two types of radiomics in a single feature vector which is then leveraged by a classifier \cite{paul2016deep, lao2017deep, chen2016automatic}. These approaches extract HCR and DLR features independently, without guaranteeing complementarity between the two sets of features. As a result, the extracted DLR may be highly redundant with the HCR, limiting the value of their combination. 

Given this context, we propose to extract DLR features that will complement the information already contained in the HCR. Our contributions are two-fold:
\begin{itemize}
    \item A deep learning method, based on the VAE framework~\cite{kingma2014}, that extracts non-redundant DLR features with respect to a predetermined set of HCR. This is achieved by minimizing the mutual information between the two types of radiomics during the training of the VAE. The resulting HCR and DLR features are leveraged to predict early markers of cancer.
    \item Validation of the proposed approach in the case of pancreatic cancer, using $2319$ training and $1094$ test subjects collected from 9 medical institutions with a split performed at the institution level. This is all the more important as most combination approaches have been solely evaluated in a cross-validation setting on mono-centric data~\cite{huynh2016digital, liu2017pulmonary, antropova2017deep}.
\end{itemize}

\section{Method}
\label{sec:method}
\begin{figure}[t]
\centering
\includegraphics[width=\textwidth]{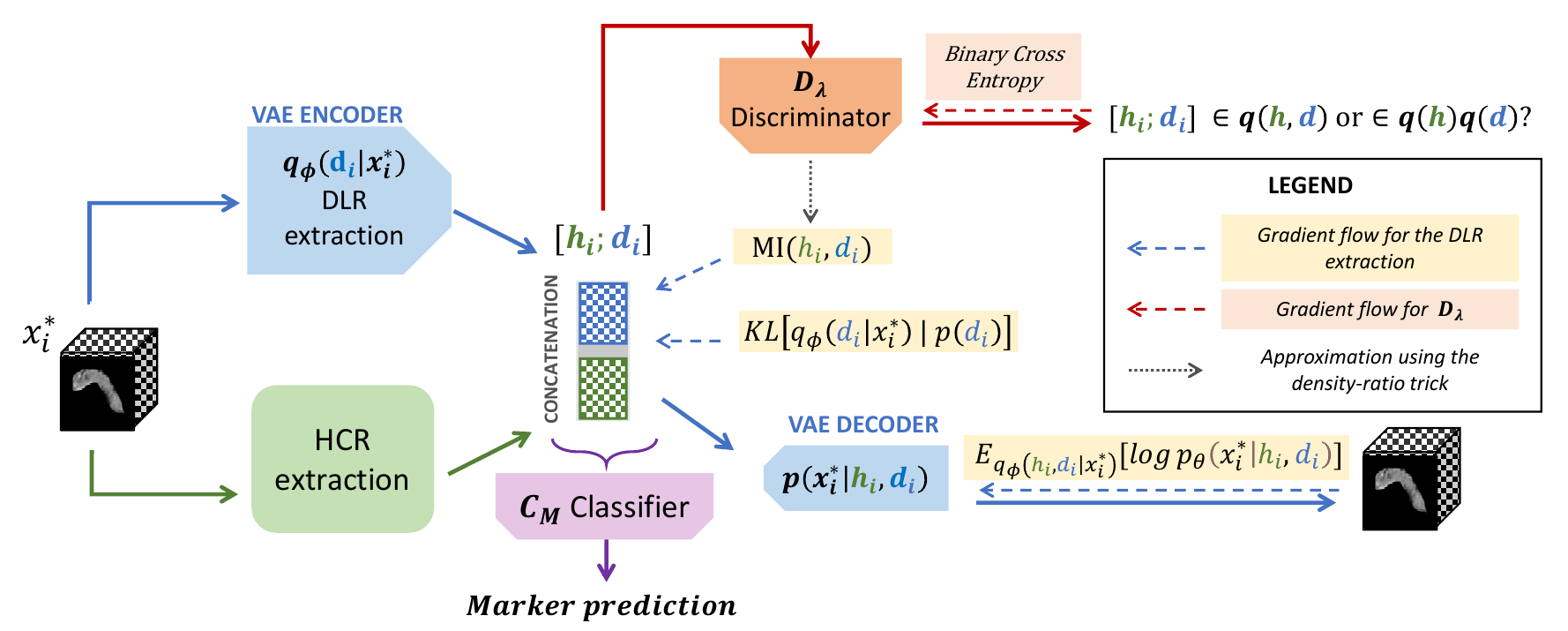}
\caption{\textbf{Overview of our method}. Starting from a masked image, Hand-Crafted Radiomics (HCR) are calculated analytically, while Deep Learning Radiomics (DLR) are extracted by the encoder of a VAE. These two types of radiomics are subsequently combined and given to the decoder for image reconstruction. The independence of HCR and DLR is enforced by the minimization of the Mutual Information (MI). The latter is approximated by the density-ratio trick~\cite{kim2018disentangling}, involving a discriminator $\mathcal{D}_{\lambda}$. Following the training of the VAE, a classifier $\mathcal{C}_{M}$ can be trained using both the HCR and DLR features to predict a specific marker of interest.} \label{fig:vae}
\end{figure}

Our method, illustrated in Figure~\ref{fig:vae}, relies on a generative model that recreates a 3D input image from the concatenation of HCR and DLR features. Feature extraction is done analytically for the HCR and through a VAE encoder for the DLR. Independence between the features is encouraged through the minimization of their mutual information, which is estimated by a discriminator relying on the density-ratio trick~\cite{kim2018disentangling}. Finally, the resulting features are given to a classifier for cancer marker prediction. 

\indent\textbf{Generative framework.}
Let $x \in \mathbb{R}^{V}$ be a 3D image acquired via a standard imaging technique, and $y \in \{0, 1\}^{V}$ the corresponding binary segmentation mask of a given organ, with $V$ the number of voxels. In order to focus on a specific organ and facilitate the extraction of specific features, we work on the masked image $x^{*} = x \times y$. We postulate the existence of a generative model enabling us to create an image $x^{*}$ from a low-dimensional representation space $[h, d]$ where $h \in \mathrm{R}^{N_h}$ and $d \in \mathrm{R}^{N_d}$ represent the HCR and DLR features with $N_h$ and $N_d$ being the number of hand-crafted and deep features, respectively. Assuming that $x^{*}$ follows an independent and identically distributed Gaussian distribution, and that $f_{\theta}$ is a non-linear function mapping the concatenation of vectors $[h, d]$ to the masked image $x^{*}$, we hypothesize the following generative process:
\begin{align}
    p_{\theta}(x^{*} \mid y, h, d) = \prod_{v=1 / y_{v} = 1}^{V}\frac{1}{\sqrt{2\pi \sigma^{2}}}\exp{\frac{(x^{*}_{v} - f_{\theta}([h, d])_{v})^{2}}{2\sigma^{2}}}
\end{align}

\indent\textbf{HCR and DLR features computation.}
We place ourselves within the VAE framework~\cite{kingma2014} and assume that $p(d)$ follows a Gaussian distribution with zero mean and identity covariance. HCR features are calculated analytically, while DLR features are computed by introducing the approximate posterior distribution $q_{\phi}(d \mid x^{*})$. We hypothesize  $q_{\phi}(d \mid x^{*}) \sim \mathcal{N}(\mu_{\phi}(x^{*}), \sigma^2_{\phi}(x^{*})\mathbf{I})$, and maximize a lower bound of the marginal log-likelihood $\log p_{\theta}(x^{*} \mid  y)$. We obtain the following loss function:
\begin{align}
\mathcal{L}_\mathrm{VAE} =  -\E_{q_{\phi}(d|x^{*})}[\log (p_{\theta}(x^{*} \mid  y, h, d)) ] + KL[q_{\phi}(d \mid x^{*}) \mid p(d)]  
\end{align}
where KL refers to the Kullback-Leibler divergence.

\indent\textbf{Mutual Information Minimization.}
To promote the independence between HCR and DLR features, we propose to minimize their Mutual Information (MI), expressed here as $KL[q(h,d) \mid q(h)q(d)]$, where $q(h,d)$ represents the joint distribution of the DLR and HCR features, and $q(h)q(d)$ the product of their marginal distributions. These terms involve mixtures with a large number of components, making them intractable. Moreover, obtaining the direct Monte Carlo estimate necessitates processing the entire dataset in a single pass. Thus, we sample from these distributions to compute the MI: to sample from $q(h, d)$, we randomly choose an image $x^{*}_{i}$, extract its HCR features $h_{i}$ as well as its DLR features $d_{i}$ using the VAE encoder, and concatenate them. Samples from $q(h)q(d)$ are obtained by concatenating vectors $h_{k}$ and $d_{j}$ with $k \neq j$. Finally, to compute the MI, we need to compute the density-ratio between $q(h, d)$ and $q(h)q(d)$. To do so, we resort to the density-ratio trick~\cite{kim2018disentangling}, which consists in introducing a discriminator $\mathcal{D}_{\lambda}([h,d])$ able to discriminate between samples from $q(h, d)$ and samples from $q(h)q(d)$. Thus, we obtain:
\begin{align}
    KL[q(h,d) \mid q(h)q(d)] = \E_{q(h,d)}\biggr[\log \frac{q(h,d)}{q(h)q(d)}\biggr] \approx \sum_i \mathrm{ReLU} \biggr(\biggr[\log \frac{\mathcal{D}_{\lambda}(h_i,d_i)}{1 - \mathcal{D}_{\lambda}(h_i,d_i)}\biggr] \biggr).
\end{align}
where the $\mathrm{ReLU}$ function forces the estimate of the MI to be positive, which prevents from back-propagating wrong estimates of the density-ratio. $\mathcal{D}_{\lambda}$ implementation is detailed in Section \ref{ssec:appendix_mi} of the appendix. 

\indent\textbf{Optimization.}
The final loss function is: 
\begin{align}
    \mathcal{L} = \mathcal{L}_\mathrm{VAE} + \kappa KL[q(h, d) \mid q(h)q(d)]
\end{align}
This loss function is composed of two terms: the left-hand term, which is the common VAE loss function and promotes the reconstruction of the masked image while regularizing the approximate posterior distribution; and the right-hand term which minimizes the MI between $q(h, d)$ and $q(h)q(d)$, and enforces the extraction of DLR features which are not redundant with HCR features. The importance of the MI in the loss function is weighted by $\kappa$, which we empirically set to $1$ (see Section~\ref{ssec:appendix_kappa} of the appendix for more details). To ensure that the density-ratio is well-estimated, as explained in \cite{kim2018disentangling}, we opt for an alternate optimization scheme between the VAE model and the discriminator $\mathcal{D}_{\lambda}$: every 5 epochs, we freeze the optimization of the VAE, train the discriminator for 150 epochs, and continue the optimization of the VAE model. 

\indent\textbf{Early cancer markers prediction.} Once the VAE model is trained, DLR can be extracted and leveraged to predict cancer markers. We propose to train, for each marker of interest, a classifier $\mathcal{C}_{M}$ based on the concatenation of HCR and DLR extracted by our model. Unlike VAE training, which is unsupervised and task-agnostic, $\mathcal{C}_{M}$ training is supervised and specific to a cancer marker.

\section{Experiments}
\label{sec:experiments}
\begin{figure}[t]
\centerline{\includegraphics[width=1.2\textwidth]{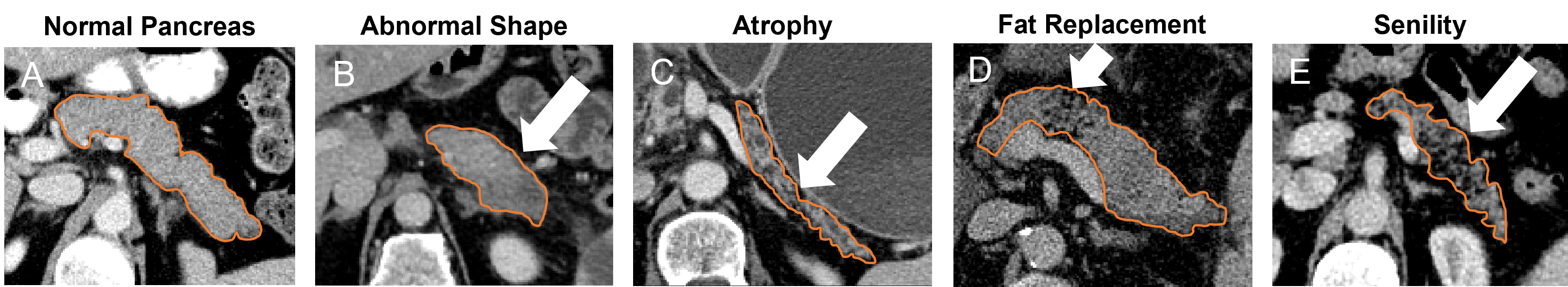}}
\caption{\textbf{Portal CT scans showing early markers of pancreatic cancer.} Pancreas are delineated in orange. (A) shows a normal pancreas. White arrows indicate an abnormal enlarged tail (B), a parenchymal atrophy (C), fat replacement in the neck of the pancreas (D) and senile characteristics (E).} \label{fig:illustrations}
\end{figure}

We illustrate our method on the pancreas, for which we aim to predict four early markers of abnormality that manifest prior to the onset of visible lesions: 
\begin{enumerate}[label=(\roman*)]
    \item \textit{Abnormal shape:} Changes in the shape of the pancreas can be associated with pancreatic cancer as the tumor growth can lead to various structural changes in the pancreas~\cite{liu2019jointshape, vetil2022learning};
    \item \textit{Atrophy:} Pancreatic atrophy may signal pancreatic cancer~\cite{miura2020focal} and can indicate small isodense lesions~\cite{yamao2020partial};
    \item \textit{Fat replacement:} Fat replacement is characterized by the accumulation of fat within the pancreas and is associated with various metabolic diseases, pancreatitis, pancreatic cancer, and precancer~\cite{khoury2017clinical, majumder2017fatty, miura2020focal}. While this mainly modifies the texture, severe fat replacement can also affect the shape by inducing lobulated margins;
    \item \textit{Senility:} Anatomical changes in the pancreas, such as pancreatic atrophy, fatty replacement and fibrosis have been documented in elderly individuals and increase the susceptibility of individuals to pancreatic cancer~\cite{khoury2017clinical, matsuda2019age}.
\end{enumerate}
These early signs are illustrated in Figure~\ref{fig:illustrations}.

\indent\textbf{Dataset.} Data were obtained from our private cohort and split into two independent datasets $\mathcal{D}^{Train}$ and $\mathcal{D}^{Test}$, containing 2319 and 1094 abdominal portal CT scans from six and three independent medical centers, respectively. The reference labels regarding the early markers previously described were obtained based on the assessment of the CT scan by a pool of 7 radiologists. Reference labels were collected for $676$ cases of $\mathcal{D}^{Train}$ and all the subjects from $\mathcal{D}^{Test}$.

\indent\textbf{Preprocessing.} 
For all the subjects, pancreas segmentation masks were obtained using a segmentation model derived from the nnU-Net~\cite{isensee2021nnunet} and manually reviewed by radiologists. The CT images and corresponding masks were resampled to $1 \times 1 \times 2$ mm$^3$ in the $(x, y, z)$ directions, and centered in a volume of size $192 \times 128 \times 64$ voxels. Images intensities were clipped to the $[0.5, 99.5]$ percentiles and standardized based on the percentiles, mean and standard deviation of the pancreas intensities in $\mathcal{D}^{Train}$.

\indent\textbf{Extracting HCR and DLR.} 32 HCR features were extracted utilizing the pyradiomics library~\cite{van2017computational}, focusing exclusively on shape and first-order intensity features (see Section~\ref{ssec:appendix_hcr} in the appendix for the comprehensive list). Complementary DLR features were extracted using the VAE model of Section~\ref{sec:method}. The architecture followed the U-Net \cite{ronneberger2015} encoder-decoder scheme without skip connections. The number of convolutional layers and the convolutional blocks were automatically inferred thanks to the nnU-Net self-configuring procedure~\cite{isensee2021nnunet} (see Section~\ref{ssec:appendix_architecture} in the appendix for details). The model was trained on $\mathcal{D}^{Train}$ for $1000$ epochs. The dimension of DLR features $d$ was set to 32, resulting in a final latent space dimension for the VAE of 64. Data augmentation consisting of rotation and cropping was applied during training.

\indent\textbf{Predicting early cancer markers.} For each marker, a logistic regression was trained based on the concatenation of HCR and DLR features extracted from the subjects in $\mathcal{D}^{Train}$ for whom reference labels were available. The logistic regression was regularized using $\mathcal{L}_2$ penalty, with a default regularization coefficient of 1. Final predictions for $\mathcal{D}^{Test}$ were derived by ensembling models obtained through a four-fold cross-validation setup.

\newcommand{\Ht}{$\mathrm{H}_{32}$}
\newcommand{\Dt}{$\mathrm{D}_{32}$}
\newcommand{\DtMi}{$\mathrm{D}_{32}^\mathrm{MI}$}

\newcommand{\Hs}{$\mathrm{H}_{64}$}
\newcommand{\HD}{$\mathrm{HD}_{64}$}
\newcommand{\HDMi}{$\mathrm{HD}_{64}^\mathrm{MI}$}

\section{Results}
\label{sec:results}

\indent\textbf{Quantitative results.}
To demonstrate the usefulness of extracting DLR with MI minimization, two VAEs were trained. Both followed the same procedure (detailed in Figure~\ref{fig:vae}) but differed only in the presence or absence of the MI minimization term in their loss function. Then, several logistic regression models with different inputs were trained in order to assess the effect of combining HCR and DLR features. In total, the following experiments were run:
\begin{itemize}
    \item \textbf{HCR only: {\Ht} and {\Hs}}. These two experiments use the 32 basic HCR features described in Section~\ref{ssec:appendix_hcr} of the appendix, and {\Hs} uses a further 32 HCR gray-level features calculated by the pyradiomics library~\cite{van2017computational} and selected by recursive feature elimination.
    \item \textbf{DLR only: {\DtMi} and {\Dt}}. 32 DLR features extracted by a VAE with and without MI minimization, respectively;
    \item \textbf{HCR + DLR: {\HDMi} and {\HD}}. 32 basic HCR features $+$ 32 DLR features extracted by a VAE with and without MI minimization, respectively.
\end{itemize}
Thus, the logistic regressions of {\Ht}, {\Dt} and {\DtMi} used vectors of size 32, while those of {\Hs}, {\HD} and {\HDMi} used vectors of size 64. Prediction results for each of the four cancer markers are presented in Table~\ref{tab:auc_results}.

\begin{table}[h]
\centering
\caption{\textbf{Pancreatic cancer marker prediction.} For each experiment, we report the means and standard deviations of the AUC (in \%) obtained by bootstrapping with $10000$ repetitions. For each line, first and second best results are in bold and underlined, respectively. The last row shows the difference in AUC compared with {\Ht}, averaged over the different markers. DLR and HCR refer to Deep Learning Radiomics and Hand-Crafted Radiomics, respectively.}
\label{tab:auc_results}
\begin{tabular}{ccc||cc||cc}
\multicolumn{1}{l}{} & \multicolumn{2}{c||}{\textbf{HCR only}}                    & \multicolumn{2}{c||}{\textbf{DLR only}}           & \multicolumn{2}{c}{\textbf{HCR + DLR}}                    \\
 & {\Ht} & {\Hs} & {\Dt} & {\DtMi} & {\HD} & {\HDMi} \\ \cline{2-7} 
\scriptsize{Abnormal Shape}       & 68.38\scriptsize{$\pm$0.07}          & 68.11\scriptsize{$\pm$0.07}          & 67.66\scriptsize{$\pm$0.07} & \textbf{72.41\scriptsize{$\pm$0.07}} & \underline{71.2\scriptsize{$\pm$0.07}}  & 70.07\scriptsize{$\pm$0.07}          \\
\scriptsize{Atrophy}              & 81.05\scriptsize{$\pm$0.06}          & \underline{81.57\scriptsize{$\pm$0.05}} & 74.08\scriptsize{$\pm$0.07} & 79.08\scriptsize{$\pm$0.06}          & 80.82\scriptsize{$\pm$0.06}          & \textbf{82.57\scriptsize{$\pm$0.06}} \\
\scriptsize{Fat Replacement}      & \underline{70.55\scriptsize{$\pm$0.07}} & 69.78\scriptsize{$\pm$0.08}          & 65.96\scriptsize{$\pm$0.08} & 65.74\scriptsize{$\pm$0.07}          & 69.28\scriptsize{$\pm$0.08}          & \textbf{71.05\scriptsize{$\pm$0.07}} \\
\scriptsize{Senility}             & 71.63\scriptsize{$\pm$0.08}          & 70.21\scriptsize{$\pm$0.08}          & 70.18\scriptsize{$\pm$0.07} & 69.1\scriptsize{$\pm$0.08}           & \underline{72.28\scriptsize{$\pm$0.08}} & \textbf{72.44\scriptsize{$\pm$0.07}} \\
\scriptsize{$\delta$ w.r.t {\Ht}}       & \textbf{-}                  & -0.48\scriptsize{$\pm$0.07} & -3.43\scriptsize{$\pm$0.07} & -1.32\scriptsize{$\pm$0.07} & \underline{0.49}\scriptsize{$\pm$0.07} & \textbf{1.13\scriptsize{$\pm$0.07}}
\end{tabular}
\end{table}

The comparison between {\Ht} and {\Hs} showed that adding 32 gray-level HCR features was not beneficial as results were similar, or even decreased: for instance, for senility, the AUC went from $71.63$ \% ({\Ht}) to $70.21$ \% ({\Hs}). On average, the AUC of {\Hs} lost -0.48 points compared with {\Ht}. These experiments demonstrated the power of the 32 basic HCR features, and the need to find complementary features that would add value.

Then, for almost all markers, {\Ht} outperformed {\Dt} and {\DtMi}, meaning that no VAE, whether trained with or without MI minimization, managed to automatically extract 32 DLR features as informative as the 32 basic HCR features used by {\Ht}. For texture-related markers, such as fat replacement and senility, MI minimization did not produce clear differences. On the other hand, on shape-related markers, the DLR features learned by {\DtMi} were shown to be more relevant than those learned by {\Dt} with a basic VAE. Thus, on average, DLR features were better when extracted by a VAE trained with MI minimization, but still proved less informative than HCR features.

Finally, experiments {\HD} and {\HDMi} showed that combining the two types of radiomics is beneficial since the average AUC gained 0.49 ({\HD}) and 1.13 \% ({\HDMi}) compared to {\Ht}. Yet, results demonstrated that minimizing the redundancy produced the best results compared with all other approaches. Indeed, in {\HD}, adding 32 DLR features produced variable results depending on the markers: compared to {\Ht}, the AUC increased by a maximum of 2.82\% for abnormal shape prediction, and dropped by a maximum of 1.27\% for predicting fat replacement. On the other hand, {\HDMi} outperformed {\Ht} on all prediction problems, meaning that the non-redundant DLR features systematically provided useful information.

\indent\textbf{Influence of the latent space.}
To explore the influence of the latent space dimension on the prediction performances, we replicated the {\HDMi} experiment with increasing size $L$ of the latent space, and reported prediction results in Table~\ref{tab:latent_size}. Table~\ref{tab:latent_size} shows that increasing the latent space size resulted in lower classification performances. Specifically, a latent space size of 32 provided the most relevant DLR features.

\begin{table}[h] 
\centering
\caption{\textbf{Pancreatic cancer marker prediction with varying latent space size.} For each experiment, a VAE with Mutual Information (MI) minimization and latent space size $L$ was trained. Predictions were obtained after training logistic regressions on 32 basic HCR features $+$ $L$ DLR features extracted by a VAE with MI minimization.  We report the means and standard deviations of the AUC (in \%) obtained on the test set by bootstrapping with $10000$ repetitions. For each line, first best results are in bold. DLR and HCR refer to Deep Learning Radiomics and Hand-Crafted Radiomics, respectively.}
\label{tab:latent_size}
\begin{tabular}{ccccccc}
\multicolumn{1}{c}{}                       & \multicolumn{1}{c}{$L=32$}         & \multicolumn{1}{c}{$L=64$}         & \multicolumn{1}{c}{$L=256$}        & \multicolumn{1}{c}{$L=512$}        & \multicolumn{1}{c}{$L=1024$}               & \multicolumn{1}{c}{$L=2048$}       \\ \cline{2-7} 
\scriptsize{Abnormal Shape}                           & \textbf{70.07\scriptsize{$\pm$0.07}}        & 69.02\scriptsize{$\pm$0.07}                 & 68.87\scriptsize{$\pm$0.07}                 & 69.91\scriptsize{$\pm$0.07}                 & 69.33\scriptsize{$\pm$0.07}                         & 68.68\scriptsize{$\pm$0.07}                 \\
\scriptsize{Atrophy}                                    & 82.57\scriptsize{$\pm$0.06}                 & 82.28\scriptsize{$\pm$0.05}                 & 81.77\scriptsize{$\pm$0.06}                 & \textbf{82.68\scriptsize{$\pm$0.05}}        & 80.9\scriptsize{$\pm$0.06}                          & 80.21\scriptsize{$\pm$0.06}                 \\
\scriptsize{Fat Replacement}                            & \textbf{71.05\scriptsize{$\pm$0.07}}        & 70.91\scriptsize{$\pm$0.07}                 & 70.23\scriptsize{$\pm$0.08}                 & 70.45\scriptsize{$\pm$0.08}                 & 69.55\scriptsize{$\pm$0.07}                         & 68.96\scriptsize{$\pm$0.08}                 \\
\scriptsize{Senility}                                   & \textbf{72.44\scriptsize{$\pm$0.07}}        & 72.02\scriptsize{$\pm$0.07}                 & 70.38\scriptsize{$\pm$0.08}                 & 71.65\scriptsize{$\pm$0.08}                 & 72.03\scriptsize{$\pm$0.07}                         & 69.6\scriptsize{$\pm$0.08}                  \\
\end{tabular}
\end{table}

\begin{figure}[t]
\centerline{\includegraphics[width=1.2\textwidth]{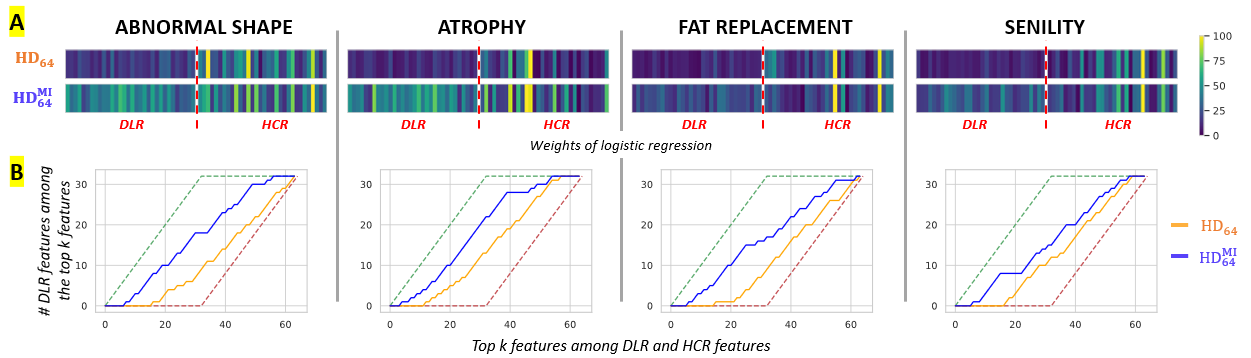}}
\caption{\textbf{Qualitative assessment of the Deep Learning Radiomics (DLR) and Hand-Crafted Radiomics (HCR) features through the coefficients of the logistic regressions.} \textit{A: Absolute value of the coefficients of the logistic regressions.} We plot, for each logistic regression corresponding to one marker, the absolute value of the coefficient for each of the 64 features. The first 32 features corresponded to DLR, while the 32 remaining features corresponded to HCR. \textit{B: Number of DLR features among the top $k$ features.} Dashed lines represent the extreme scenarios in which all 32 DLR are more informative than all 32 HCR (green) or all 32 HCR are more informative than all 32 DLR (red).} \label{fig:lr_coef}
\end{figure}

\indent\textbf{Qualitative results.}
To visualize the effect of the extracted DLR features, we looked at the absolute value of the logistic regression weights for {\Dt} and {\DtMi} in two ways. In Figure~\ref{fig:lr_coef}-A, the absolute value of these coefficients are displayed. The higher the absolute value of the coefficient, the higher its importance in the logistic regression prediction. When the MI was not minimized, HCR features had stronger importance than DLR ones. On the other hand, when we encouraged the independence between the two types of features through MI minimization, the contribution of DLR features to the prediction increased. Figure~\ref{fig:lr_coef}-B shows the number of DLR features among the $k$ features with highest importance, for increasing values of $k$. {\HDMi} and {\HD} are shown in blue and orange, respectively. In addition, two extreme scenarios are shown: one where the logistic regression is predominantly influenced by the DLR features (in green), and another one where the logistic regression is primarily driven by the HCR features (in red). We can see that the blue curve approached the green curve, meaning that DLR features from {\HDMi} contributed more to the outcome prediction. When the MI was not minimized, DLR features had less influence on the predictions as the orange curve approached the scenario in which DLR would be ignored.

\indent\textbf{Reconstruction performances.}
To explore the reconstruction performances of the VAE, we computed the average $l_2$ error per voxel between the original test images and their corresponding reconstructions. Upon applying nnU-Net's~\cite{isensee2021nnunet} automatic intensity normalization procedure, voxel intensities were observed to range from $-3$ to $2.3$. Specifically, we employed a VAE with a latent space dimension of $L=32$ and MI minimization during training. The resulting reconstruction error was found to be $(4.4\pm1.4) \times 10^{-3}$, which was comparable to the $l_2$ error obtained from a VAE trained without MI minimization, amounting to $(4.1\pm1.4) \times 10^{-3}$. These observations suggest that the introduction of MI minimization did not significantly impact the quality of the reconstructed images, neither resulting in deterioration nor improvement. Additionally, Table~\ref{tab:l2_error} further explores the relationship between reconstruction performance and latent space sizes, demonstrating that increasing the latent space size did not have a discernible effect on the quality of the reconstructions.
\begin{table}[h]
\centering
\caption{\textbf{Reconstruction performances with varying latent space sizes.} For each experiment, a VAE with Mutual Information minimization and latent space size $L$ was trained. We report the $l_2$ error per voxel between the original image and its reconstruction, with voxel intensities varying in $[-3, 2.3]$.}
\label{tab:l2_error}
\begin{tabular}{c|c|c|c|c|c|c}
\multicolumn{1}{c|}{}                       & \multicolumn{1}{c|}{$L=32$}         & \multicolumn{1}{c|}{$L=64$}         & \multicolumn{1}{c|}{$L=256$}        & \multicolumn{1}{c|}{$L=512$}        & \multicolumn{1}{c|}{$L=1024$}               & \multicolumn{1}{c}{$L=2048$}       \\ \cline{2-7} 
$l_2$ error $\times 10^3$ & 4.4\scriptsize{$\pm$1.4} & 4.4\scriptsize{$\pm$1.4} & 4.4\scriptsize{$\pm$1.4} & 4.4\scriptsize{$\pm$1.4} & 4.3\scriptsize{$\pm$1.4} & 4.3\scriptsize{$\pm$1.5}
\end{tabular}
\end{table}

\section{Discussion and conclusion}
\label{sec:conclusion}
We presented a method to learn DLR features that are not redundant with HCR ones. The method was based on the well-known VAE framework~\cite{kingma2014} that extracted DLR features from masked images in an unsupervised manner. The complementarity between the two types of radiomics features was enforced by minimizing their MI, and the resulting features were used to train classifiers predicting different cancer markers. Experiments in the case of four early markers of pancreatic cancer indicated that our method increased prediction performances with respect to two state-of-the-art approaches. These findings suggest that our approach holds potential to improve patient survival outcomes. Qualitative results confirmed the advantages of minimizing the MI during training, as it resulted in the generation of DLR features that were complementary to HCR features and more prominently utilized for marker prediction. These results were obtained on a large and independent test set, which is particularly important as radiomics models require robust validation strategies to ensure their generalization and reproducibility when applied to new datasets~\cite{aerts2014decoding}. With this in mind, it might be interesting to further encourage this feature efficiency by imposing independence between the DLR features themselves. Another research avenue could be to simplify the proposed pipeline by developing an end-to-end network capable of performing both feature extraction and classification tasks within a unified framework. Achieving this objective would necessitate the simultaneous training of the feature extractor and multiple sub-networks for each classification task. However, this approach might pose challenges in terms of training complexity, particularly due to the presence of substantial class imbalances across the various classification tasks. Alternatively, another possibility is to train an end-to-end convolutional neural network (CNN). Although more direct in nature, this approach would entail the training of a separate CNN for each question, which could be computationally heavier compared to the calibration of a logistic regression based on a single feature extractor, as suggested in our current work. Future studies should also address the interpretability of the extracted DLR features, as this aspect was not covered in the present work.

\bigskip

\indent\textbf{Acknowledgments} This work was partly funded by a CIFRE grant from ANRT \# $2020/1448$.

\section{Appendix}
\subsection{Estimating the Mutual Information}
\label{ssec:appendix_mi}

\begin{figure}[h]
\centering
\includegraphics[width=\textwidth]{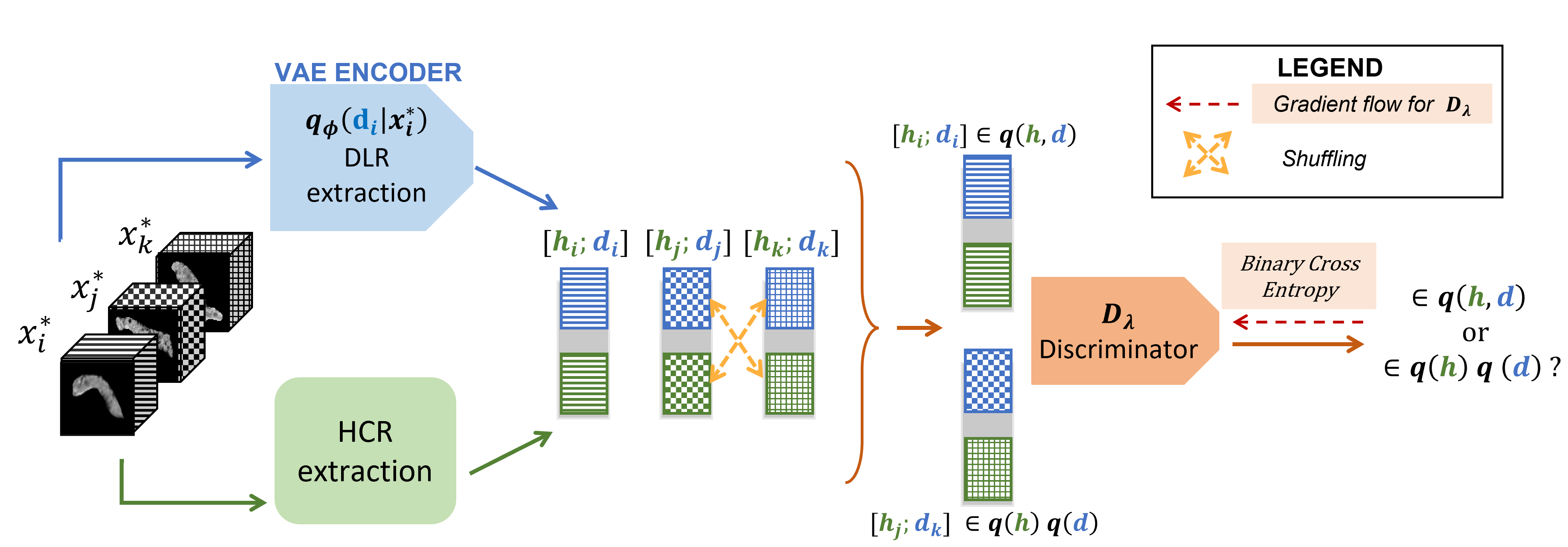}
\caption{\textbf{Training $\mathcal{D}_{\lambda}$.} Given three different input images $x^{*}_{i}$, $x^{*}_{j}$ and $x^{*}_{k}$, the corresponding HCR and DLR features are computed: $h_{j}$, $h_{j}$, $h_{k}$ and $d_{i}$, $d_{j}$, $d_{k}$. Samples from $q(h, d)$ are obtained by concatenating features of a same image ($h_i$ and $d_i$ for instance), while samples from $q(h)q(d)$ are obtained by concatenating $h_{k}$ and $d_{j}$ with $k \neq j$.} \label{fig:mi}
\end{figure}

The Mutual Information (MI) is estimated following the density-ratio trick~\cite{kim2018disentangling} which requires to train a discriminator $\mathcal{D}_{\lambda}$ predicting whether concatenated radiomics vectors $[h, d]$ come from $q(h,d)$ or $q(h)q(d)$. Samples for training $\mathcal{D}_{\lambda}$ are obtained following the procedure shown in Figure~\ref{fig:mi}. In practice, $\mathcal{D}_{\lambda}$ is modeled as a 2-layer Multi Layer Perceptron with ReLu activation, which is trained by minimizing a binary cross-entropy (BCE) loss term. Once the discriminator is trained, the MI between HCR and DLR features can be approximated as follows:

\begin{align}
    \mathrm{MI}(h, d) = \E_{q(h,d)}\biggr[\log \frac{q(h,d)}{q(h)q(d)}\biggr] \approx \sum_i \mathrm{ReLU} \biggr(\biggr[\log \frac{\mathcal{D}_{\lambda}(h_i,d_i)}{1 - \mathcal{D}_{\lambda}(h_i,d_i)}\biggr] \biggr).
\end{align}

\subsection{Influence of the hyperparameter $\kappa$}
\label{ssec:appendix_kappa}

The final loss function for training our model is: 
\begin{align}
    \mathcal{L} = \mathcal{L}_\mathrm{VAE} + \kappa  KL[q(h, d) \mid q(h)q(d)]
\end{align}
where $\kappa$ is a hyperparameter weighting the importance of the the mutual information in the total loss function. Table~\ref{tab:kappa} reports prediction results obtained with different values of $\kappa$. According to these results, $\kappa$ was set to $1$ in all our experiments.

\begin{table}[h]
\centering
\caption{\textbf{Cancer marker prediction scores for different values of $\kappa$.} For each experiment, we report the means and standard deviations of the AUC (in \%) obtained by bootstrapping with $10000$ repetitions. For each line, best result is in bold.}
\label{tab:kappa}
\begin{tabular}{ccccc}
                         & \textbf{$\kappa = 0.01$}     & \textbf{$\kappa = 0.1$} & \textbf{$\kappa =1$}         & \textbf{$\kappa = 10$}         \\ \cline{2-5} 
\textbf{General Shape}   & 70.44\scriptsize{$\pm$0.07}         & 70.01\scriptsize{$\pm$0.07}     & 70.07\scriptsize{$\pm$0.07}          & \textbf{71.03\scriptsize{$\pm$0.07}} \\
\textbf{Atrophy}         & 80.82\scriptsize{$\pm$0.05}          & 81.43\scriptsize{$\pm$0.06}     & \textbf{82.57\scriptsize{$\pm$0.06}} & 80.77\scriptsize{$\pm$0.06}          \\
\textbf{Fat Replacement} & 69.52\scriptsize{$\pm$0.08}          & 70.5\scriptsize{$\pm$0.07}      & \textbf{71.05\scriptsize{$\pm$0.07}} & 68.65\scriptsize{$\pm$0.08}          \\
\textbf{Senility}        & \textbf{73.14\scriptsize{$\pm$0.08}} & 72.36\scriptsize{$\pm$0.08}     & 72.44\scriptsize{$\pm$0.07}          & 72.38\scriptsize{$\pm$0.08}         
\end{tabular}
\end{table}

\subsection{HCR features extraction}
\label{ssec:appendix_hcr}

32 HCR features were extracted using the pyradiomics library~\cite{van2017computational}:
\begin{itemize}
    \item \textbf{14 shape features} describing the size and shape of the pancreas
    \begin{itemize}
        \item Mesh Volume
        \item Voxel Volume
        \item Surface Area
        \item Surface Area to Volume ratio
        \item Sphericity
        \item Maximum 3D diameter
        \item Maximum 2D diameter in the axial plane
        \item Maximum 2D diameter in the coronal plane
        \item Maximum 2D diameter in the sagittal plane
        \item Major Axis Length
        \item Minor Axis Length
        \item Least Axis Length
        \item Elongation
        \item Flatness   
    \end{itemize}
    \item \textbf{18 first-order intensity features} describing the intensities distribution within the organ
    \begin{itemize}
        \item Energy
        \item Total Energy
        \item Entropy
        \item Minimum
        \item $10^{th}$ percentile
        \item $90^{th}$ percentile
        \item Maximum
        \item Mean
        \item Median
        \item Interquartile Range
        \item Range
        \item Mean Absolute Deviation
        \item Robust Mean Absolute Deviation
        \item Root Mean Squared
        \item Skewness
        \item Kurtosis
        \item Variance
        \item Uniformity
    \end{itemize}
\end{itemize}
More details about each feature can be found \href{https://pyradiomics.readthedocs.io/en/latest/features.html}{on the online documentation}.

\subsection{Model Architecture}
\label{ssec:appendix_architecture}

\begin{figure}[h]
\centering
\includegraphics[width=\textwidth]{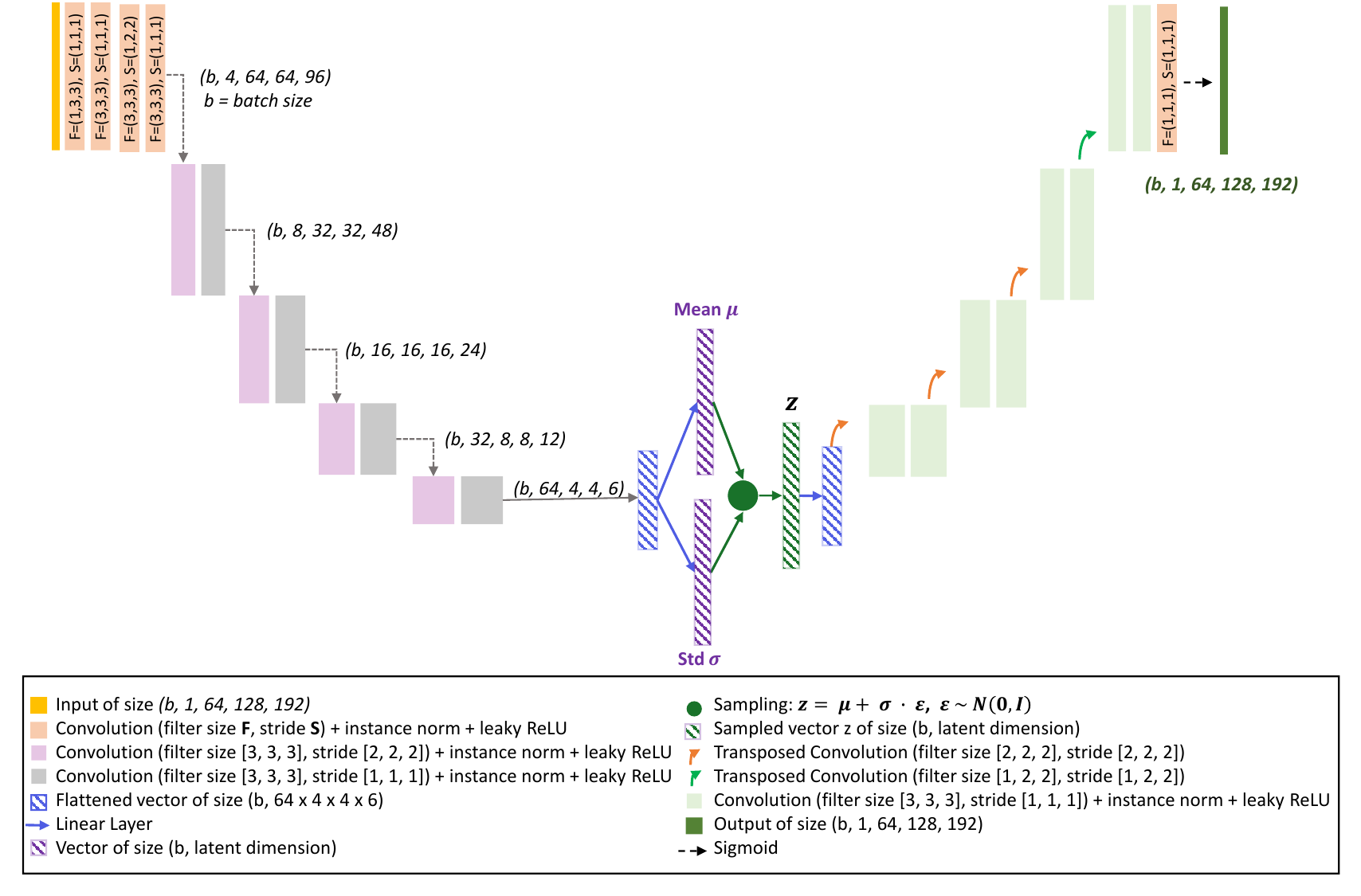}
\caption{\textbf{Architecture of the proposed VAE}} \label{fig:architecture}
\end{figure}

As detailed in Figure \ref{fig:architecture}, the proposed variational autoencoder (VAE) followed a 3D encoder-decoder architecture. The network topology (number of convolutions per block, filter sizes) was chosen based on the nnU-Net self-configuring procedure~\cite{isensee2021nnunet}, resulting in $1,110,240$ trainable parameters. The VAE was trained on $1000$ epochs with a batch size of size $32$. Every 5 epochs, the VAE was frozen and the discriminator $\mathcal{D}_{\lambda}$ was trained for 150 epochs with a batch size equal to the total training dataset. The VAE and $\mathcal{D}_{\lambda}$ were optimized using two independent Adam optimizers with a learning rate of $10^{-3}$. 

%
%
%
\bibliographystyle{splncs04}
\bibliography{biblio}

\end{document}